\documentclass[twocolumn,aps,prl]{revtex4}
\usepackage{graphicx}
\newcommand{\eins}{\mbox{$1 \hspace{-1.0mm}  {\bf l}$}}
\newcommand{\one}{\mbox{$1 \hspace{-1.0mm}  {\bf l}$}}
\newcommand{\be}{\begin{equation}}
\newcommand{\ee}{\end{equation}}
\newcommand{\bea}{\begin{eqnarray}}
\newcommand{\eea}{\end{eqnarray}}
\newcommand{\half}{\mbox{$\textstyle \frac{1}{2}$}}

\newcommand{\shalf}{\mbox{$\textstyle \frac{1}{\sqrt{2}}$}}
\newcommand{\sthird}{\mbox{$\textstyle \frac{1}{\sqrt{3}}$}}

\newcommand{\ket}[1]{ | \, #1  \rangle}
\newcommand{\bra}[1]{ \langle #1 \,  |}

\newcommand{\proj}[1]{\ket{#1}\bra{#1}}

\newcommand{\calp}{\mbox{$\cal P$}}
\newcommand{\mytext}[1]{\mbox{ #1}}

\newcommand{\Tr}{ {\rm Tr}}

\begin{document} 

\title{Reflections upon separability and distillability}

\author{Dagmar Bru\ss$^1$, J. Ignacio Cirac$^2$, Pawe\l\ Horodecki$^3$,
Florian Hulpke$^1$, Barbara Kraus$^2$, Maciej Lewenstein$^1$, and Anna
Sanpera$^1$}
\address{$^1$Institut f\"ur Theoretische
Physik,  Universit\"at Hannover, D-30167 Hannover, Germany\\
$^2$Institut f\"ur Theoretische Physik, Universit\"at Innsbruck,
 A-6020 Innsbruck, Austria\\
$^3$Faculty of Applied Physics and Mathematics,\\
Technical University of Gda\'nsk, 80--952 Gda\'nsk, Poland}

\begin{abstract}
We present an abstract formulation  of the so-called Innsbruck-Hannover
programme  that investigates  quantum correlations and entanglement in 
terms of convex sets. We present a unified description of optimal
decompositions of quantum states and the optimization of witness operators
that detect whether a given state belongs to a given convex set. We
illustrate the abstract formulation with several examples, and discuss
relations between optimal entanglement witnesses and
$n$-copy non-distillable states with non-positive partial
transpose. 
\end{abstract}
\maketitle

\section{I. Introduction}
The characterization and classification of entangled states, introduced by
Schr\"odinger
\cite{Sch:35},  is perhaps {\em the most challenging open problem of
modern quantum theory}. There are at least four important issues that
motivate us to study the entanglement problem:

I. Interpretational and philosophical motivation: Entanglement plays an
essential role in apparent ``paradoxes'' and counter-intuitive consequences 
of quantum mechanics \cite{EPR:35,Sch:35,Zurek}.
 
II. Fundamental physical motivation: 
The characterization of entanglement is one of the most fundamental 
open problems of quantum mechanics. It should answer the question what the 
nature of quantum correlations in composite systems \cite{Per:95} is.

III. Applied physical motivation: 
Entanglement plays an essential role in applications of quantum 
mechanics to 
quantum information processing (quantum computers \cite{qcomp}, 
quantum cryptography \cite{qcrip} and quantum communication \cite{qcomm}). 
The resources needed to  
implement a particular protocol of quantum information processing  are 
closely linked to the entanglement properties of the states used  in the 
protocol. In particular, entanglement lies at the 
heart of quantum computing.

IV. Fundamental mathematical motivation: 
The entanglement problem is directly related to one of the most challenging 
open problems of linear algebra and functional analysis: the
characterization and classification of positive maps on ${\cal C}^*$ 
algebras \cite{posmaps1,jami,posmaps2,posmaps3}.

In the recent years there have been several excellent reviews treating the
entanglement problem, i.e.  the question whether a given state is
entangled and if it is, how much entangled it is. Considerable effort has
been also devoted to the distillability problem, i.e. the question
whether one can distill a maximally
entangled state from many copies of a given state,
by means of local operations and classical
communication.
Entanglement is discussed in the context of quantum communication in
\cite{hororev}. B. Terhal summarizes the use of witness operators for
detecting entanglement in \cite{terhal}.  Various entanglement measures
are presented in the context of the theorem of their uniqueness
in \cite{mdonald}. Some of us have recently written a ``primer'' \cite{primer},
which aims at introducing the non-expert reader 
to the problem of separability and distillability of quantum states, an
even more elementary ``tutorial'' \cite{tutor}, and a
 partial ``tutorial'' \cite{tutor1}.

The present paper, which contains material of invited lectures given by
Dagmar Bru\ss\ and Maciej Lewenstein at the 
Second Conference on ``Quantum Information Theory and
Quantum Computing'',
held by the  European Science Foundation in Gda\'nsk
in July 2001, is not yet another review or tutorial. It is addressed to
experts in the area and reports the recent progress in the applications of
the so-called Innsbruck-Hannover (I-Ha) programme of investigations 
of quantum correlations and entanglement (see Fig. 1 for the logo of the
programme). The new results of the present paper are as follows. First, we
present an abstract formulation of the I--Ha programme  in  terms of convex
sets. Second, we present a unified description of optimal decompositions
of quantum states  and optimization of witness operators that detect
whether a given state belongs to a given convex set. We illustrate the
abstract formulation with several examples, and point out analogies and
differences between them. Finally, we discuss relations between
optimal entanglement witnesses and
$n$-copy non-distillable states with non-positive partial transpose.


\begin{figure}
\centering
\includegraphics[width=7cm]{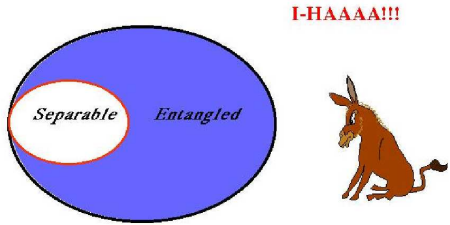}
\caption{\small Logo of the I--Ha programme.}
\label{logo}
\end{figure}

The paper is organized as follows. In section II we briefly
present the milestones in the investigations of the  separability
problem. Then, in section III we present the abstract formulation of
the I-Ha programme, and discuss such concepts as optimal decompositions,
edge states,   witness operators and their optimization, and 
canonical  forms of witnesses and linear maps.
Several applications are presented in section IV; they concern the
questions of  separability of quantum states, Schmidt number of quantum
states and classification of mixed states in three-qubit systems. Furthermore 
we discuss relations between optimal witnesses
and non-distillable  states with non-positive partial transpose (NPPT
states)\cite{barbara}. We conclude in section V. 

\section{II. Milestones in the separability problem} 

The separability problem can be formulated in simple words as follows:
Given a  physical state in a many-party system described by a density
operator, is it separable, i.e. can it be prepared by local actions and
classical  communication,  or not? Mathematically, this question
reduces to the question whether the state $\varrho$ can be written as 
convex combination of projectors onto product states; for two-party
systems  a mixed state $\varrho$ is called {\em separable}
   iff it can be written as \cite{wer}
 \be
 \varrho \, { =}\, \sum_ip_i\proj{e_i}\otimes\proj{f_i}\ ,
 \label{sep}
 \ee 
 otherwise it is {\em entangled}. Here the coefficients $p_i$ are
 probabilities, i.e. $0\leq p_i \leq 1$ and $\sum_i p_i=1$.
 Note that in general 
neither $\{|e_i \rangle \}$ 
nor $\{|f_i \rangle \}$ have to be orthogonal.

It is relatively easy to detect whether a given pure state is entangled
or not, since only the simple tensor product states are separable.
An appropriate tool to investigate this in  two-party systems 
is provided by the Schmidt decomposition \cite{Per:95}. It is also known that
entangled pure states incorporate genuine quantum correlations, can be
distilled and violate some kind of Bell's inequalities \cite{????}.
The situation is completely different for mixed states, where up to now we
cannot answer the question of separability of a given state in general.

For the perspective of the present paper, we  can point out the 
following  milestones in the investigations of the separability 
problem \cite{foot}:

\begin{itemize}
\item The mathematical definition of separable states 
and its suprising consequences for Bell inequalities
has been given by Reinhard Werner in 1989 \cite{wer}.

\item Following  the mixed states teleportation idea 
\cite{Popescu0} the first scheme of how to improve entanglement
of a spatially mixed state is provided
by Sandu Popescu \cite{Popescu} 
(significantly extended by Nicolas Gisin 
with the help of the POVM technique \cite{gisin}).

\item
In 1996 the IBM group \cite{pur} has given first examples of 
purification and distillation procedures 
and related them to quantum error correcting codes
(applied by the Oxford group to quantum privacy amplification 
\cite{oxf}).

\item In 1996 Asher Peres \cite{peres} has formulated the necessary
condition  for separability that requires positivity of the partial
transpose.

\item In the same year 1996 it has been proven
that the  PPT condition  is also sufficient in $2\otimes 2$ and
$2\otimes 3$ systems in Ref. \cite{ppt}. In the same paper 
the relation of the separability problem to the
theory of positive maps has been rigorously formulated.

\item
First examples of entangled states with positive partial
transpose in  higher dimensions (PPTES) have been
found in \cite{pawel}. Here a
necessary condition for separability based on the analysis of the range 
of the state in  question  (the so-called ``range'' criterion)
has been formulated as well.

\item Since 1998  PPT
entanglement has been identified in a series of papers
as non--distillable, but sometimes activable;
this type of entanglement has been termed ``bound entanglement"
\cite{horodist,horobe,horoact,reduction}.

\item In 1998  the I-Ha programme was initiated with the first paper
on optimal decompositions \cite{M&A}.

\item Several families of PPTES's were constructed by the IBM group,
using a systematic procedure based on unextendible product
bases (UPBs) \cite{UPB}.

\item Barbara Terhal presented in 2000 the first construction of
non-decomposable entanglement witnesses (nd-EW) and non-decomposable
positive maps (nd-PM) related to UPBs \cite{witness}.

\item In 2000 the existence of non--distillable states with non--positive
partial transpose (non--distillable NPPT states) has been conjectured by
us \cite{npt1}  and by the IBM group \cite{npt2}.

\end{itemize}

The I--Ha programme  initially concentrated on the separability
problem for two-party systems. Only recently its scope has enlarged and
incorporated studies of multi-party systems, continuous variables (CV)
systems, and systems of identical particles. This has led us to an
abstract general formulation of the I-Ha programme  which we present in
the next section. In this and the following section we provide the necessary
references when we discuss concrete applications of the
I--Ha programme.  

\section{III. Innsbruck--Hannover programme}

This section is divided into several subsections that describe the
abstract formulation of the I--Ha programme, as well as  successive
elements of the programme: optimal decompositions, edge states, witness
operators and their optimization. We present here for the first time
the
optimization procedure as a form of optimal decomposition 
of witness operators and discuss
canonical forms of witnesses and corresponding linear maps.

\subsection{Abstract formulation of the problem}

The problem that we consider is schematically represented in Fig. 2. 
We consider two sets of self-adjoint  operators $S_1$ and
$S_2$ acting on a Hilbert space $H=H_A\otimes H_B\otimes\ldots$ of some
composite quantum  system. We denote the space of operators acting on
$H$ by ${\cal B}(H)$. This is a Hilbert space with the scalar product
induced by the trace operation.  Both sets $S_1$ and
$S_2$ are compact, i.e. closed and
bounded,  and convex. The set
$S_1$ is a proper subset of
$S_2$.  In many applications those sets are  subsets of the set of
physical operators, i.e. they contain trace class, positive operators of
trace one. We will consider, however, also the case when the sets $S_1$ and
$S_2$ will contain witness operators, which by definition are not positive
definite, although they are trace class and have trace one.  We will
assume that the properly defined identity operator belongs to $S_1$. For
Gaussian states in systems  with continuous variables the states will be
represented by their correlation matrices, and the sets $S_1$ and
$S_2$ will be the sets of correlation matrices of certain properties. In
this latter case, the sets
$S_1$ and
$S_2$ will be closed and convex, but not necessarily bounded. Quite
 generally we assume that the sets are invariant under local 
invertible operations.


\begin{figure}
\centering
\includegraphics[width=7cm]{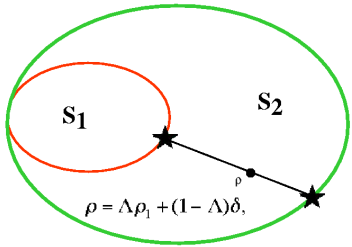}
\caption{\small Scheme of the abstract formulation of the problem. }
\label{abstract}
\end{figure}

The class of questions that we consider is: Given $\varrho\in S_2$, does it
belong to
$S_1$, or not? 

\subsection{Optimal decomposition}

A useful tool to study such questions is provided by the
optimal decomposition:
\\
\noindent{\bf Theorem 1.} Every $\varrho\in S_2$ can be decomposed as a
convex combination of an operator $\varrho_1\in S_1$ and $\delta\in
S_2\setminus S_1$,
\be
\varrho=\Lambda\varrho_1+(1-\Lambda)\delta,
\ee
where $1\ge \Lambda\ge 0$. The above decomposition is optimal in the
sense that $\delta$ can  only 
have a trivial decomposition with
the corresponding $\Lambda=0$. There exists the best decomposition, for
which $\Lambda$ is maximal and unique.
\\

The proof of the above Theorem is the same as the proof of the similar
theorem discussed in the context of separability 
and PPTES's \cite{M&A,karnas}. Note that the optimal decomposition is
not unique, and that when $\Lambda=1$ then $\varrho\in S_1$. The
coefficient $\Lambda$ corresponding to the best decomposition is unique,
but generally it is not known if the best decomposition is. So far, this
has been proven only for the best separable decompositions for $2\otimes
2$ systems in Ref. \cite{M&A} and for $M\otimes N$ systems in Ref.
\cite{karnas}. Note also that  optimal decompositions can be obtained
in a constructive way, by subtracting from $\varrho$ operators in $S_1$,
keeping the remainder in $S_2$. It is sufficient to subtract
$\varrho_1\in S_1$ that are extremal points of the convex set $S_1$. 

\subsection{Edge operators}

The operators $\delta$ that enter the optimal decompositions are called
 edge operators. They are defined as follows:
\\
\noindent{\bf Definition 1.} A state $\delta\in S_2$ is an edge operator iff
for every $\varrho_1\in S_1$ and $\epsilon>0$ the operator
$\varrho'=(1+\epsilon)\delta-\epsilon\varrho_1$ does not belong to $S_2$.
\\

Note that the edge operators  characterize fully the operators that 
belong to $S_2\setminus S_1$, and every extremal point of the convex set 
$S_2$ is  either an element of $S_1$, or an edge operator.  
The construction of examples of edge operators is possible by
constructing optimal decompositions.   It is usually possible to check
constructively if a given operator  is an edge operator.

\subsection{Witnesses and their optimization}

The existence of witness operators is a direct consequence of the
Hahn--Banach theorem \cite{rudin}. Witness operators are defined
as follows: 
\\
\noindent{\bf Definition 2.} An operator $W$ is a witness operator
(pertaining to the pair of sets $S_1$ and $S_2$) iff for every $\sigma\in
S_1$ it holds that ${\rm Tr}(\sigma W)\ge 0$, and there exists a $\varrho\in
S_2$ for which ${\rm Tr}(\varrho W)< 0$. We say that $W$ detects $\varrho$.
We normalize the witnesses demanding that ${\rm Tr}(W)=1$.
\\

For every $\varrho\in S_2\setminus S_1$ there exists  a witness operator which
detects it. Witness operators in the context of the separability problem,
i.e. entanglement witnesses have been discussed already in the seminal
paper \cite{ppt}. In the recent years first examples of the  so-called
non--decomposable entanglement witnesses were provided by B. Terhal
\cite{terhal} who introduced the name ``entanglement witness''.
Terhal's construction provides witnesses of PPT
entangled states that are obtained from UPBs.  The most general
procedure of constructing  decomposable and non-decomposable
entanglement witnesses was provided by us \cite{char,opti}.

Non-trivial entanglement witnesses are operators
that have some negative eigenvalues.
However, the positivity of mean values on product states leads to
additional non-trivial properties.
To illustrate this we shall present here the proof of some
property which was known for specialists (see
for example \cite{tutor1}) but the proof of it has nowhere 
been published explicitely.
\\
\noindent{\bf Observation.}
If $W_{A}$ and $W_{B}$ stand for partial
reductions of $W$ then the range of $W$ belongs to
tensor products of ranges of the reductions.
\\

Note that the above property 
is possessed by {\it all} bipartite states
(see \cite{bechan}) but 
{\it not} by all operators (see the operator
$|\Psi_{+} \rangle \langle \Psi_{+}|-\frac{1}{2}|0\rangle
|0\rangle \langle 0| \langle 0|$ with 
the triplet wavefunction $\ket{\Psi_{+}}$).
To prove the Observation it is only needed to show
that $P_{W_A}\otimes I W P_{W_A}\otimes I=W$, where
 $P_{W_A}$ stands for the projection
on the  support of $W_{A}$.
We have to prove
(i) $P_{W_A}^{\perp}\otimes I W P_{W_A}^{\perp} \otimes I=0$
and
(ii) $P_{W_A}^{\perp}\otimes I W P_{W_A} \otimes I=
P_{W_A}\otimes I W P_{W_A}^{\perp} \otimes I=0$.
Let $|e\rangle \in P_{W_{A}}$
and $|e\rangle^{\perp} \in P_{W_{A}}^{\perp}$.
Then by the very definition
$ \langle e^{\perp} | \langle  f | W | e^{\perp}\rangle | f \rangle \geq 0$.
The same inequality holds for all $|f'\rangle$-s that form 
an orthogonal
basis together with  $|f\rangle$ in the
 Hilbert space of the right system.
If we sum  up all of them we get
that the resulting sum is equal to
$ \langle e^{\perp} |W_{A}|e^{\perp}\rangle \geq 0$.
But the latter is zero because $W_{A}$ must be positive
(otherwise contrary to its definition $W$ would be negative on a product 
state proportional to $|e^{\perp} \rangle \langle e^{\perp}| 
\otimes I$, see \cite{tutor1}).
So the zero is every element of the sum, as it is non-negative.
This gives
$ \langle e^{\perp} | \langle  f | W 
 |e^{\perp} \rangle | f \rangle= 0$
concluding the proof of (i).

Now by the very definition
of an entanglement witness we have
for any $| e^{\perp} \rangle | f \rangle $, 
  for any $\alpha \in [0,1]$ and
$\phi\in [0,2\pi] $ that 
$(\alpha \langle e^{\perp}|+
e^{-i\phi}\sqrt{1-\alpha^{2}}) \langle e|) \langle  f | W 
(\alpha| e^{\perp}
\rangle + e^{i \phi}
\sqrt{1-\alpha^{2}}) |e \rangle) | f \rangle \geq 0$.
Taking $\phi=1,-1,\pi/2, -\pi/2$,
using (i) and taking the limit
$\alpha \rightarrow 0$ we get
that the cross terms $ \langle e^{\perp} | \langle  f | W 
| e\rangle | f \rangle $ and
$ \langle e | \langle  f | W 
|e^{\perp}  \rangle | f \rangle $
vanish, which concludes (ii) and the whole proof.
The Observation is an illustration that  entanglement witnesses
are subjected to rather strong restrictions
and have properties similar to states (another 
such property is the convexity of the 
set of witnesses). The above reasoning can be  easily extended 
to multiparticle systems. Let us now turn to the interpretation 
of entanglement witnesses.

Geometrically, the meaning of the witnesses is explained in Fig. 3. If
$\varrho$ does not belong to the convex and compact set $S_1$, then there 
exists a hyperplane that separates $\varrho$ from $S_1$. For each such
hyperplane in the space of operators, there exists a corresponding $W$.
>From the figure it is clear that the most useful witnesses are
those that correspond to hyperplanes that are as tangent as possible
to the set $S_1$. It is thus useful to search for such witnesses, i.e.
to optimize witness operators. 


\begin{figure}
\centering
\includegraphics[width=7cm]{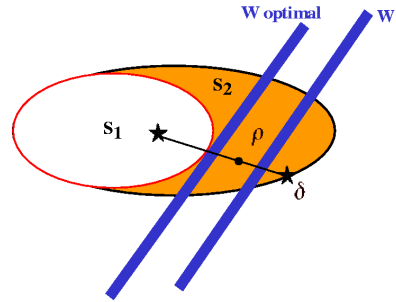}
\caption{\small Schematic representation of a witness operator 
and an optimal witness operator.}
\label{swiadki}
\end{figure}

To this aim one introduces the notion of a witness being {\it finer} 
than another one
\cite{opti}. A witness $W_2$ is called finer than $W_1$, if every
operator that is detected by $W_1$ is also detected by $W_2$. A witness
$W$ is optimal if there does not exist a witness 
that is finer than $W$. Let
us denote by    $P_1$ the set of operators $Z_1$ such that
for every $\varrho_1\in S_1$, ${\rm Tr}(Z_1\varrho_1)\ge0$. The set $P_1$ is
convex and for most of the applications that we consider compact. It contains
witness operators and the so-called 
{\em pre-witness} operators, that have
positive trace with the operators from $S_1$ but do not  detect anything. 
Similarly, we define the set $P_2$ of operators $Z_2$ such that
for every $\varrho \in S_2$, ${\rm Tr}(Z_2\varrho_2)\ge0$. The set  $P_2$ is
also convex and (in most applications) compact, and contained in $P_1$.
We have the following theorems:
\\
\noindent{\bf Theorem 2.} If a witness $W_2$ is finer than $W_1$, then 
$W_1=(1-\epsilon)W_2+\epsilon Z_2$, where $Z_2\in P_2$.
\\
The proof is a straightforward generalization of the proof of Lemma 2
from Ref. \cite{opti}. Similarly, we obtain the generalization of 
Theorem 1 in Ref. \cite{opti} (see also \cite{hulpke}):
\\
\noindent{\bf Theorem 2'.} A witness $W$ is optimal iff for all $Z_2\in
P_2$ and $\epsilon >0$, $W'=(1-\epsilon)W-\epsilon Z_2$ is not a
witness, i.e. it does not fulfill ${\rm Tr}(Z_1\varrho_1)\ge0$ for all
$Z_1$ from $P_1$.
\\
>From the above Theorem 2' we see that  optimal witnesses can be
regarded as edge states (pertaining to the pair $P_2\subset P_1$); we will
call them {\em edge witnesses}. We have a straightforward generalization of 
Theorem 1:
\\
\noindent{\bf Theorem 3.} Every witness $W\in P_1$ can be decomposed as
a convex combination of an operator $Z_2\in P_2$ and an
optimized witness $W_{opt}\in P_1\setminus P_2$,
\be
W=\Lambda Z_2+(1-\Lambda)W_{opt},
\ee
where $1\ge \Lambda\ge 0$. The above decomposition is optimal in the
sense that $W_{opt}$ can  only have a trivial 
decomposition with
the corresponding $\Lambda=0$. There exists the best decomposition, for
which $\Lambda$ is maximal and unique.
\\

Note that, as above,  the optimal decomposition is
 not unique, and that when $\Lambda=1$ then $W\in P_2$ and $W$
ceases to be a witness, since it does not detect anything.  Note also
that  optimal decompositions can be obtained in a constructive way, by
subtracting operators $Z_2 \in P_2$ from $W$, keeping the remainder in
$P_1$. It is sufficient to subtract
$Z_2\in P_2$ that are extremal points of the convex set $P_2$. 

Finally, note that optimal witnesses (i.e.  edge witnesses)
characterize fully the witness operators that  belong to $P_1\setminus
P_2$, and every extremal point of the convex set 
$P_1$ is  either an element of $P_2$, or an edge witness.  
The construction of  examples of edge witnesses is possible by
constructing optimal decompositions.   

\subsection{Canonical form of witnesses and linear maps}
 
It is relatively easy to formulate the generalization of the theorems that
were proven in Ref. \cite{char}. In particular, we have the 
following theorem about the canonical form of the witness operators. 
Let $\hat{\eins}=\eins/{\rm Tr}\eins$, then we have:
\\
\noindent{\bf Theorem 4.} Every witness operator has the canonical form
\be 
W=(1+\epsilon)Z_2-\epsilon \hat{\eins},
\ee
where $Z_2\in P_2$, and there exists an edge operator $\delta$ such
that ${\rm Tr}(Z_2\delta)=0$. The parameter $\epsilon$ fulfills
\be
\epsilon/(1+\epsilon)\le \inf_{\varrho_1\in S_1}{\rm Tr}(Z_2\varrho_1).
\ee
\\
The proof that $\epsilon>0$ is the following. We consider the family 
$W+\tilde \epsilon \hat{\eins}$ for increasing $\tilde \epsilon$. For some 
finite range of $\tilde \epsilon$'s,
 $0\le \tilde \epsilon\le\epsilon_0$, this family consists of 
witnesses (which are positive on all elements of $S_1$, and are 
non--positive on at least one element of $S_2$). Let us consider $\epsilon_0$  
for which ${\rm Tr}(W+\epsilon_0 \hat{\eins})\varrho$ 
is non-negative for all $\varrho\in S_2$. From compactness of $S_2$ there 
exists 
a 
$\varrho_0$ such that  ${\rm Tr}(W+\epsilon_0\hat{\eins})\varrho_0=0$. There 
cannot exist 
an  optimal decomposition of $\varrho_0=\Lambda\varrho_1+(1-\Lambda)\delta$ 
with 
$\Lambda\ne 0$, since otherwise ${\rm Tr}(W+\tilde \epsilon\hat{\eins})
\varrho_1$ 
would be 
negative for $\tilde \epsilon<\epsilon_0$, {\it ergo} 
$\varrho_0=\delta$ is an edge operator. From this argument it follows 
immediately that $(1+\epsilon_0)Z_2=W+\epsilon_0\hat{\eins}$, and thus that
${\rm Tr}(Z_2\varrho_1)$ is strictly positive for all $\varrho_1\in S_1$. 
 Due to compactness the same has to hold for the infimum of this quantity.

There exists an elegant isomorphism \cite{jami} between operators and
linear maps,  which allows to construct maps that ``detect'' states that
do not belong  to $S_1$. We will consider here the isomorphism for 
two-party systems,  but it can be generalized and is in fact quite useful to
study  many-party systems \cite{homanypar}.  

Let an operator $O_{BC}\in {\cal B}(H_B\otimes H_C)$. Then we can define
a map
\be
E: \ {\cal B}(H_B)\to {\cal B}(H_C),
\ee
such that for any $\varrho_B\in{\cal B}(H_B)$, we have
\be
E(\varrho_B)={\rm Tr}_B(O_{BC}\varrho_B^{T_B})\in {\cal B}(H_C),
\ee
where the trace is taken over the space $H_B$ only, and $T_B$ denotes the
transposition in the space $H_B$. 

Note that if we denote by $H_{B'}$ a space isomorphic to $H_B$, we then 
have
\be
\eins_{B'}\otimes E(|\Psi\rangle\langle \Psi|)=O_{B'C},
\ee
where 
\be
|\Psi\rangle = \sum_{i=1}^{{\rm dim}H_B}|i\rangle_{B'}|i\rangle_B/
\sqrt{{\rm dim}H_B}
\ee
is the maximally entangled state. Note that if $O_{B'C}$ is a witness that
detects $\varrho_{B'C}$ (i.e. if ${\rm Tr}_{B'C}(O_{B'C}\varrho_{B'C})<0$),
then the adjoint map 
\be
\eins_{B'}\otimes E^{\dag}(\varrho_{B'C})= {\rm
Tr}_C(O_{BC}^{T_B}\varrho_{B'C})= \varrho_{B'B}\in{\cal B}(H_{B'}\otimes H_B) 
\ee
transforms $\varrho_{B'C}$ into an operator $\varrho_{B'B}$ which is not
positive definite, and ${\rm Tr}_{B'B}(\varrho_{B'B}|\Psi\rangle\langle
\Psi|)<0$. If the adjoint map acting on some $\varrho$  produces a
non-positive operator, we say that it detects $\varrho$. Maps detect more
operators than witnesses. For instance, if the map detects $\varrho_{B'C}$,
then it also detects $O_{B'}\varrho_{B'C}O_{B'}^{\dag}$ for any $O_{B'}$,
where
$O_{B'}$ is an element of the linear group $GL(H_{B'})$. 

If the operator $O_{BC}$ has some properties, then it implies some
properties of the corresponding map $E[O_{BC}]$. For instance, if $O_{BC}$
is positive definite, then  $E[O_{BC}]$ is a completely positive 
map (CPM), if
$O_{BC}$ is an entanglement witness, then  $E[O_{BC}]$ is  positive (PM),
if $O_{BC}$
is a decomposable  entanglement witness, i.e. $O_{BC}=P+Q^{T_B}$, where
both $P,Q\in {\cal B}(H_B\otimes H_C)$ are positive, 
then 
$E[O_{BC}]$ is  a decomposable positive map,
i.e. a convex sum of a CPM and another CPM composed with the partial
transposition in $H_B$. Similarly, if $O_{BC}$
is a non--decomposable  entanglement witness, i.e. $O_{BC}$
cannot be represented as $P+Q^{T_B}$, where both $P,Q\in {\cal
B}(H_B\otimes H_C)$ are positive,  then 
$E[O_{BC}]$ is  a non--decomposable positive map. Schmidt number witnesses
\cite{adm} are related to the so-called $k$--positive maps \cite{barpaw}.

In all of the above mentioned examples, the maps have some positivity
properties on the operators from the corresponding set $S_1$ (separable
state, PPT states, states of Schmidt number $k$ etc). It is not clear if
that will always be the case, if one considers a general abstract formulation
of the problem. Quite generally, however, we can say that if $W_{BC}$ is
a witness, i.e. ${\rm Tr}(W_{BC}\sigma_{BC})\ge 0$ for every $\sigma_{BC}$
from
$S_1$, then the corresponding adjoint map 
\be
\eins_{B'}\otimes E^{\dag}(\sigma_{B'C})= {\rm
Tr}_C(W_{BC}^{T_B}\sigma_{B'C})= \sigma_{B'B} 
\ee 
has the property that
$\langle\Psi|\sigma_{B'B}|\Psi\rangle\ge 0$.

The analysis of the maps completes our presentation of the I--Ha
programme. Note that the knowledge of the canonical form of witnesses
allows to construct the canonical form of the corresponding maps. 

\section{IV. Applications of the I--Ha Programme}

In this section we will discuss in more detail 
the application of the I-Ha
programme to concrete problems: the separability problem, the relation
between optimal separability witnesses and non-distillable NPPT states, 
the Schmidt number of mixed states, and finally the classification of 
mixed states of three-qubit systems. 

Before turning to a more detailed discussion let us list the results 
and applications of the I-Ha programme achieved so far:

\begin{itemize} 

\item The optimal decomposition in the context of separability (optimal
separable approximations and  best separable approximations, BSA) have
been introduced in Ref.
\cite{M&A}. The theory of optimal decompositions, and in particular of 
optimal decompositions of PPT states is presented in
\cite{karnas}. Important results concerning the best separable
approximations for $2\otimes 2$ systems have been obtained by 
B.-G. Englert \cite{englert}. An analytic construction of the BSA in this case
was recently achieved by Ku\'s and Wellens \cite{kus}.

\item The construction of ``edge'' states and the solution of the
separability problem for low rank states were
discussed by us in Refs. \cite{pra1,pra2,pra3} for $2\otimes N$, $M\otimes N$ and
$2\otimes 2\otimes N$ systems, respectively. 

\item The canonical form of witnesses and positive maps and their
optimization is discussed in Refs. \cite{char,opti,adm}.

\item The I--Ha programme  has also been applied to study quantum
 correlations in fermionic systems \cite{fermi}, 
Schmidt number witnesses \cite{adm}, and three-qubit 
systems \cite{abls}.

\item The programme  is presented partially in the primer \cite{primer}
and in the tutorial \cite{tutor}.

\item The programme has been  also applied to study systems with
continuous variables (CV). We have provided 
the first example of PPT states for CV
systems \cite{pptcv}, and solved the separability problem for two-party 
Gaussian states with an arbitrary number of modes \cite{gauss1},
and for three-party systems each having a single mode \cite{gauss2}.
Werner and Wolf \cite{werwol} used the I-Ha approach to provide first
examples of the Gaussian PPT states -- ``edge" states in our terminology. 

\item The I-Ha programme was extended to study separability and entanglement of
quantum operations \cite{oper}.

\end{itemize}

We will now discuss some applications, and in particular we will show
how they fit into the general scheme discussed in the previous
section. 
\subsection{Application I - the separability problem}

Two kinds of applications were discussed by us in this context: the
application to the general problem of separability, in which
 $S_1$ is the set of separable states $S$, whereas $S_2$ is the set of
all states, denoted here as $B$ \cite{M&A,karnas,opti}. Another application
concerns the problem of distinguishing separable states from PPT
entangled states: in this case $S_1=S$, whereas $S_2$ is the set of PPT
states, so that the set of PPTES's is given by $PPT\setminus S_1$. For
the first application we have:

\noindent{\bf Proposition 1.}\cite{M&A} Every state $\rho\in {\cal
B}(H_A\otimes H_B)$ can be decomposed as
\be
\varrho=\Lambda\sigma+(1-\Lambda)\delta,
\ee
where $\sigma\in S$, and $\delta$ is an edge state, which in this
situation means a state whose range $R(\delta)$ does not contain any
product vector $|e,f\rangle$. 

Similarly, for 
the second scenario we have: 

\noindent{\bf Proposition 2.}\cite{pra1,karnas} Every state $\varrho\in
{\cal B}(H_A\otimes H_B)$ having the PPT property can be decomposed as
\be
\varrho=\Lambda\sigma+(1-\Lambda)\delta,
\ee
where $\sigma\in S$,  and $\delta$ is a PPT edge state, which in this
situation means a state whose range $R(\delta)$ does not contain any
product vector $|e,f\rangle$, such that the range of the partially
transposed state $R(\delta^{T_A})$ contains at the same time the partially
complex conjugated vector $|e^*,f\rangle$. 

Note that a similar proposition can be proven for the case when
$S_1=PPT$ and $S_2=B$. Such an optimal decomposition is difficult to realise
in a constructive way, since the subtraction of a projector onto a product
vector
must be replaced in this case by the subtraction of a PPT state 
$\sigma$  whose
range fulfills at the same time  $R(\sigma)\subset R(\delta)$,
$R(\sigma^{T_A})\subset R(\delta^{T_A})$.

Let us denote by $r(\varrho)$ the rank of $\varrho$. The studies of  PPT edge
states has led us to the following proposition (\cite{pra1,pra2}, see
\cite{pra3} for the generalization to 
$2\otimes 2\otimes N$ systems):

\noindent{\bf Proposition 3.} If a state $\varrho$ acting in an
$M\otimes N$ dimensional Hilbert space fulfills that
\be
r(\varrho)+r(\varrho^{T_A})\le 2MN-M-N+2
\ee
then there exists generically a finite number of product states that
belong to  $R(\delta)$, such that the range of the
partially transposed state $R(\delta^{T_A})$ contains at the same time
the partially complex conjugated vector $|e^*,f\rangle$. 

Note that  in such a case it is easy to check if $\varrho$  is separable,
since we know the finite number of projectors onto product vectors that
can enter the decomposition  of $\varrho$ into a convex sum of projectors
on product vectors. This means that for $r(\varrho)+r(\varrho^{T_A})\le
2MN-M-N+2$ the separability problem is essentially solved. 

It is useful to quote at this point the propositions that provide the
explicit form of entanglement
 witnesses (EW) for  edge states and  canonical forms of
witnesses for the two cases  introduced above,
$S_2=B$ and $S_2=PPT$: 

\noindent{\bf Proposition 4.} Given an edge state $\delta$, 
\be
W=P-\epsilon \eins,
\ee
where $P\ge 0$, $R(P)=K(\delta)$,  and
$$\epsilon\le \inf_{|e,f\rangle}\langle e,f|P|e,f\rangle$$
is a decomposable EW that detects $\delta$. 

\noindent{\bf Proposition 4'.} Given a PPT edge state $\delta$, 
\be
W=P+Q^{T_A}-\epsilon \eins,
\ee
where $P\ge 0$, $R(P)=K(\delta)$, $Q\ge 0$, $R(Q)= K(\delta^{T_A})$,  and
$$\epsilon\le \inf_{|e,f\rangle}\langle e,f|P+Q^{T_A}|e,f\rangle$$
is a non-decomposable EW (nd-EW) that detects $\delta$.

Conversely,

\noindent{\bf Proposition 5.} If $W$ is a decomposable EW then
there exists an edge state $\delta$ such that 
\be
W=P-\epsilon \eins,
\ee
where $P\ge 0$, $R(P)=K(\delta)$,   and
$$\epsilon\le \inf_{|e,f\rangle}\langle e,f|P|e,f\rangle.$$ 

\noindent{\bf Proposition 5'.} If $W$ is a non-decomposable EW, then
there exists a PPT edge state $\delta$ such that 
\be
W=P+Q^{T_A}-\epsilon \eins,
\ee
where $P\ge 0$, $R(P)=K(\delta)$, $Q\ge 0$, $R(Q)= K(\delta^{T_A})$,  and
$$\epsilon\le \inf_{|e,f\rangle}\langle e,f|P+Q^{T_A}|e,f\rangle.$$ 

Note that proposition 4(4') and 5(5') together with 
the optimization allow to
formulate necessary conditions for witness operators to be extremal in
the convex set of all witnesses. The Jamio\l kowski isomorphism translates
those conditions into necessary conditions for positive maps to 
be extremal \cite{char}.

The two kinds of optimization (related to pairs of sets $S,B$ and $S,PPT$)
are schematically illustrated in Fig. 4. It is perhaps easier to
understand them when one looks at Fig. 5. Here we see that the set 
of $pre-EW$
of decomposable EW together with B is convex and compact (this is a set of
operators $W$ that fulfill the condition ${\rm Tr}(W\sigma)\ge 0$ for all
$\sigma\in PPT$). Similarly, adding the set of non-decomposable
witnesses to this set, one obtains yet another convex and compact set 
of $pre-nd-EW$
which contains operators $W$ such that ${\rm Tr}(W\sigma)\ge 0$ for all
$\sigma\in S$.   
\begin{figure}
\centering
\includegraphics[width=7cm]{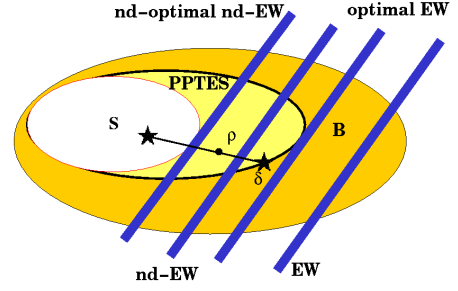}
\caption{\small Schematic representation of a  decomposable witness
operator, an optimal decomposable witness operator,
a non--decomposable witness operator, and  optimal non--decomposable
witness operator (see hyperplanes from right to left).}
\label{swiadkient}
\end{figure}
\begin{figure}
\centering
\includegraphics[width=7cm]{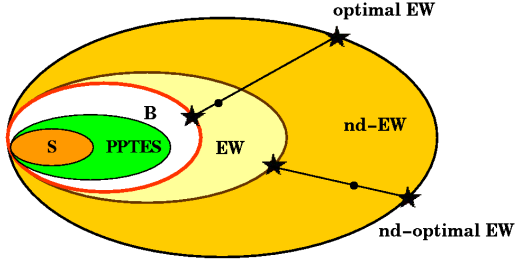}
\caption{\small Schematic representation 
of the optimization of decomposable or non--decomposable entanglement
witnesses as search for  optimal decompositions and  decomposable, or
non-decomposable ``edge'' witnesses.}
\label{swiadkient2}
\end{figure}

Figure 5 allows us immediately to write for the case $S_1=B$ and
$S_2=pre-EW$:
\\
\noindent{\bf Proposition 6.} Every $EW\in pre-EW$ can be decomposed as a
convex combination of an operator $\varrho\in B$ and 
$\delta\in pre-EW\setminus B$,
\be
EW=\Lambda \varrho +(1-\Lambda)\delta,
\ee
where $1\ge \Lambda\ge 0$. In the above decomposition $\delta$ is an 
optimal entanglement witness in the sense that $\delta$ 
can 
only have a trivial decomposition with the corresponding $\Lambda=0$. There
exists the best decomposition, for which $\Lambda$ is maximal and unique.
\\
Similarly, for non-decomposable EW's we have:
\\
\noindent{\bf Proposition 7.} Every $nd-EW\in pre-EW$ can be decomposed
as a convex combination of an operator $\varrho\in pre-EW$ and $\delta\in
pre-nd-EW\setminus pre-EW$,
\be
EW=\Lambda \varrho +(1-\Lambda)\delta,
\ee
where $1\ge \Lambda\ge 0$. In the above decomposition $\delta$ is an 
optimal nd-entanglement witness in the sense that 
$\delta$ can 
only have a trivial decomposition with the corresponding $\Lambda=0$. There
exists the best decomposition, for which $\Lambda$ is maximal and unique.
\\
The proofs are the same as in the original works
(c.f. \cite{M&A,pra1,karnas}).

\subsection{Application II -- Optimal witnesses and non-distillable NPPT 
states} 

Let us first remind the readers that it has been recently conjectured 
\cite{npt1,npt2} that there exist states with non-positive partial transpose 
that are non-distillable. In particular, in Ref. \cite{npt1} we have formulated
the following
{\bf Conjecture.}
Let $\dim H_A=\dim H_B=d$, and let
$$\varrho_W(\beta) = \frac{1}{N_W} (\calp_s + \alpha(\beta) \calp_a),$$
\begin{equation}
\varrho_W^{T_A}(\beta)=\frac{1}{\tilde{N}_W} (\eins - \beta \calp),
\end{equation}
where $\calp_{s,a}$ are the projectors onto the symmetric respectively
 antisymmetric 
subspace, $\calp$ is the projector onto a maximally entangled vector
$|\Psi \rangle= \sum_{i=1}^d
|i \rangle_A |i \rangle_B/\sqrt{d}$, 
$\alpha(\beta) =\frac{d+\beta}{d-\beta}$ and $N_W, \tilde{N}_W$ are 
normalization constants. Then, $\varrho_W(\beta)$ is a non-distillable NPPT 
state
for $\beta \in (1,d/2]$.
\\

Note that for $-1\leq\beta <1$, $\varrho_W(\beta)$ is PPT (it is in fact 
separable). The above conjecture is supported by numerical evidence for
2 and 3 copies of $\varrho_W(\beta)$, and the fact that $\varrho_W(\beta)$ is 
$n$-copy
non-distillable for $1 < \beta \leq \beta_n$. The later statement means, that 
in this case one cannot project $n$ copies of $\varrho_W(\beta)$ onto a 
$2 \otimes 2$-dimensional subspace of 
$H_A^{\otimes n} \otimes H_B^{\otimes n}$, such that the 
resulting $2 \otimes 2$ state would be distillable, i.e. would not have the PPT
property. The best bounds on $\beta_n$ are $\beta_1=d/2$, $\beta_2 = 
(2+ d)/4$, etc.
Unfortunately $\beta_n \to 1$, when $n \to \infty$.\\
It is interesting to observe that the existence of $n$-copy non-distillable 
NPPT states allows the construction of optimal EW's. To this aim we present 
the 
following:\\
{\bf Proposition 8} Let $\dim H_{A'} = \dim H_{B'} =2$, and 
let $\calp_a$ be the projector onto the singlet state $ (|0\rangle_{A'} 
|1 \rangle _{B'} - |1 \rangle_{A'} |0 \rangle_{B'})/\sqrt{2}$ in 
$H_{A'} \otimes H_{B'}$. Let $W(\beta)$ act on 
$(H_{A'} \otimes H_{A}^{\otimes n}) \otimes (H_{B'}
\otimes H_B^{\otimes n})$, where 
\be
W(\beta)= \calp_a^{T_{A'}} \otimes (\varrho_W(\beta))^{\otimes n}.
\ee
Then we have:
\begin{description}
\item [i)]
 $W(\beta)$ is an optimal non-decomposable EW for $0 <\beta \leq \beta_n$.
\item [ii)]
 $W(\beta)$ is an nd-optimal non-decomposable EW for $n=1$, $\beta =d/2$.
\end{description}
Additionally, if the conjecture is true we have:\\
{\bf Proposition 9} If the conjecture is true, the $W(\beta)$ defined in 
proposition 8 is
\begin{description}
\item [i)]
 an optimal non-decomposable EW for $\beta \in (1,d/2]$.
\item [ii)]
 an nd-optimal non-decomposable EW for $\beta=d/2$.
\end{description}
In the above propositions ``optimal EW'' means optimal with respect to 
subtraction
of positive operators ($S_1=B, S_2=\mbox{pre-EW}$). The proof of proposition 8
follows directly from the results of Ref. \cite{opti}.
\\

It is easy to see that if $|e,f\rangle$ is a product vector from
$(H_{A'} \otimes H_A^{\otimes n})\otimes 
(H_{B'} \otimes H_B^{\otimes n})$, then 
$\langle e,f | W(\beta)|e,f \rangle \geq 0$, provided 
$\langle \Psi_2|(\varrho_{W}^{T_A}(\beta))^{\otimes n}|\Psi_2\rangle \geq 0$ 
for all
$|\Psi_2\rangle \in H_A^{\otimes n} \otimes H_B^{\otimes n}$ of
Schmidt rank 2. On the other hand $W(\beta)$ is not positive, so it 
detects some states, i.e. $W(\beta)$ is an EW. Moreover,
$\langle e,f| W(\beta)|e,f\rangle=0$ for all 
$|e \rangle = |E \rangle |F\rangle_A$, $|f \rangle = |E^*\rangle |G\rangle_B$.
Since such vectors span the whole Hilbert space, $W(\beta)$ is an optimal 
EW. From Ref. \cite{opti} we know, however, that an 
optimal decomposable EW must 
be of the form
\be
W(\beta) =Q^{T_A}, \mbox{ where } Q \geq 0.
\ee
This is not the case, ergo $W(\beta)$ is an optimal non-decomposable EW for
$0 \leq \beta \leq \beta_n$, which proves part (i) of the proposition. 
Proving
part (ii) is achieved by observing that for $\beta=d/2$ there exists
another set of product vectors $|\tilde{e},\tilde{f}\rangle$ such that
$\langle \tilde{e},\tilde{f} |W^{T_A}(\frac{d}{2})|\tilde{e},\tilde{f} 
\rangle =0$,
and $|\tilde{e},\tilde{f}\rangle$ span the whole Hilbert space. The vectors 
$|\tilde{e},\tilde{f}\rangle$ are such that
\be
\Tr_{A'B'}(\calp_a |\tilde{e},\tilde{f}\rangle \langle \tilde{e},\tilde{f}|)=
|\psi_2 \rangle \langle\psi_2|\ ,
\ee
where $|\psi_2\rangle$ are vectors of Schmidt rank 2 in 
$H_A \otimes H_B$ such that 
\be
\langle \psi_2 | \varrho_W^{T_A}\left(\frac{d}{2}\right)|\psi_2 \rangle=0.
\ee
Quite generally $|\psi_2\rangle = \frac{1}{\sqrt{2}}
(|e,e^* \rangle + |e_\perp, e_\perp^* \rangle)$ where $|e \rangle$ is an 
arbitrary vector from $H_A$, and $|e_\perp\rangle$ is an arbitrary vector
from $H_A$ orthogonal to $|e\rangle$. The proof of proposition 9 is 
analogous.\\
Note that propositions 8 and 9 do not involve a specific form of
$\varrho_W(\beta)$, and are valid for all NPPT non-distillable states. 
Note also, 
that proposition 8 implies immediately the recent result in
\cite{pptdist}, that 1-copy non-distillable NPPT states are PPT-distillable.
To this aim we observe that if $W(\beta)$ is already non-decomposable 
for $n=1$, then it detects some PPT state $\sigma$, i.e.
\be
\Tr(W(\beta)\sigma) <0\ ,
\ee
that implies that
\be
\Tr_{AA'BB'}(\sigma^{T_{A'A}}[\calp_a \otimes \varrho_W^{T_A}(\beta)]) <0\ ,
\ee
which in turn allows to define a linear PPT map
$E_{W(\beta)}(\varrho) = \Tr_{AB}(\sigma^{T_A} \varrho^{T_A}) = \Tr_{AB} 
(\sigma \varrho)=
\tilde{\varrho}$
which transforms $\varrho$ into a  matrix $\tilde{\varrho}$ 
in dimension $2 \otimes 2$ which is 
NPPT, since $\Tr (\tilde{\varrho}^{T_{A'}} \calp_a) < 0$.

In \cite{barbara} we develop a general formalism which connects
entanglement witnesses to the distillation and 
activation properties of a state. There we also 
show its applications to three-party states and 
point out how it can be generalized to an arbitrary number 
of parties. 

\subsection{Application III -- Schmidt number of mixed states}

The I-Ha programme can be applied to detect how many degrees
of freedom of a given bipartite state are entangled -- this is the
interpretation of the so-called Schmidt number of a mixed
state. The Schmidt number is a generalization of the Schmidt rank
of pure bipartite states and was introduced
in \cite{barpaw} (more implicitly defined in \cite{Vidal}):
a given state $\varrho$ can be expanded as
\be
\varrho = \sum_i p_i \proj{\psi_i^{r_i}} \ ,
\ee
with $p_i\geq 0$ and $\sum_i p_i=1$.
Here $r_i$ denotes the Schmidt rank of the pure state $\ket{\psi_i}$.
The maximal Schmidt rank in this decomposition is called $r_{\rm max}$.
The Schmidt number $k$ of $\varrho$ is given as the minimum of 
$r_{\rm max}$ over all decompositions,
\be
k=\mytext{min}\{ r_{\rm max} \}\ .
\ee
Namely, $k$ tells us how many degrees of freedom have at least to be
entangled in order to create $\varrho$. This number cannot be greater than
 $M$, the smaller dimension of the two subsystems.

\begin{figure}
\centering
\includegraphics[width=7cm]{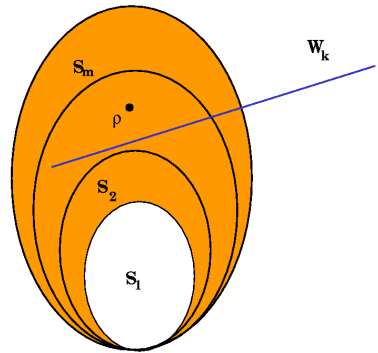}
\caption{\small Schematic representation of Schmidt classes and their
detection by Schmidt witnesses.}
\label{schmidt}
\end{figure}

With this definition one realises that the set of all states consists
of compact convex subsets of states that carry the same Schmidt number.
We denote the subset that contains states of Schmidt number
$k$ or less by $S_k$. These Schmidt classes  are successively embedded
into each other, $S_1 \subset S_2 \subset ... \subset S_M$.
This is illustrated in Figure \ref{schmidt}.

In \cite{adm} we generalized the concept of entanglement witnesses to
Schmidt number witnesses. The generalization is straightforward:
\\
\noindent{\bf Definition 3.} An operator $W_k$ is a witness operator 
of Schmidt number $k$
 iff for every $\sigma\in
S_{k-1}$ it holds that ${\rm Tr}(\sigma W_k)\ge 0$, and there exists a $\varrho\in
S_k$ for which ${\rm Tr}(\varrho W_k)< 0$. We say that $W_k$ detects $\varrho$.
We normalize the witnesses demanding that ${\rm Tr}(W_k)=1$.
\\
\\
Applying the I-Ha programme to Schmidt classes, we can write a state $\varrho$ 
as a convex combination of a state from Schmidt class $k$ and an edge
state,
\be
\varrho = \Lambda \varrho_k +(1-\Lambda) \delta \ ,
\ee
where the edge state $\delta$ has no vectors of Schmidt rank 
$\leq k$ in its range. This decomposition can be obtained by subtracting
projectors onto vectors of Schmidt rank $\leq k$ from $\varrho$,
while requiring the resulting state to be positive.

A {\em canonical} form for a  witness of the Schmidt class
$k$ is given by 
\be
W_{k} = Q -\epsilon \eins \ ,
\ee
where $Q$ is a positive operator that fulfills 
$K(Q)=R(\delta)$, for some edge state $\delta$
with Schmidt number $k$. We define
$\epsilon \equiv \mytext{inf}_{\ket{\psi}} \bra{\psi}Q\ket{\psi}$,
where $\ket{\psi}$ has Schmidt rank $\leq k-1$. This construction
guarantees that the requirements from Definition 3 are met:
the parameter $\epsilon$ is chosen such that 
 ${\rm Tr}(\sigma W_k)\ge 0$ holds for every $\sigma\in
S_{k-1}$, and $\delta$ is detected by $W_{k}$.  

The optimization of Schmidt witnesses can be performed in an analogous
way as for entanglement witnesses and is discussed in \cite{adm,hulpke}.
An example of an optimal (unnormalized) 
witness of Schmidt number $k$ in $H^M\otimes H^M$ 
is given by
\be
W_{k,opt}=\eins-\frac{M}{k-1}\calp \ ,
\label{example}
\ee
where $\calp$ is the projector onto the maximally entangled state
$|\Psi\rangle=\sum_{i=1}^{M}\ket{ii}/\sqrt{M}$. The fact that the
maximal squared overlap between $|\Psi\rangle$ and a vector with
Schmidt rank $k$ is
 $k/M$ leads to the
correct properties for a $k$-Schmidt witness.

\subsection{Application IV -- Mixed three-qubit states}

\begin{figure}
\centering
\includegraphics[width=7cm]{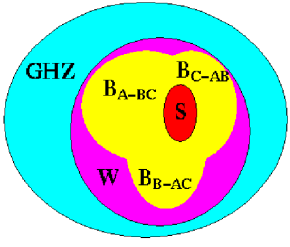}
\caption{\small  Classification of mixed three-qubit states.}
\label{three}
\end{figure}

Recently studies of tripartite states have been incorporated in the 
I-Ha programme \cite{abls}: 
composite systems of three qubits can be classified
according to Figure \ref{three}. This figure should
be regarded as  an intuitive 
picture of the structure of this set. 

The class $S$ denotes separable states, i.e. states that can be 
written as a convex combination of product states of the three
parties. The class $B$ stands for biseparable states, i.e. those
that have vectors in their decomposition in which two out of the 
three parties are entangled, and the third one is in a product state
with the two entangled ones. There are three possibilities for
such biseparable states, which are indicated as three petals in
the figure. The biseparable class is understood to be the convex
hull of them. The biseparable states are embedded in the 
$W$-class. This is the set of states that also have vectors of the
form 
\be
\ket{\psi_{W}} = \sthird (\ket{100}           +\ket{010} +\ket{001})\ 
\label{w}
\ee
in their decomposition. These so-called W-vectors were introduced
in \cite{duer}, and shown to be 
locally inequivalent to the GHZ-vectors, which
are of the form
\be
\ket{\psi_{GHZ}} =\shalf (
\ket{000}+\ket{111})\ .
\label{ghz}
\ee 
In Figure \ref{three} the GHZ-class, i.e. the set of states that have
GHZ-vectors in their decomposition, surrounds the W-states. This is the
correct ordering of states, as each of the inner sets is required to be
compact. 
By studying the most general form of a W-state and a GHZ-state, as given in
\cite{tony}, one realizes that 
swapping the role of the classes  GHZ and W would lead to a GHZ-class
that is not compact. The reason is that infinitesimally close to any W-state
there  always lies a GHZ-state.

Again, one can construct witnesses to detect the class of a given mixed
three-qubit state. We called them biseparable witnesses,
W-witnesses and GHZ-witnesses. Their definition is
a straightforward extension of Definition 3. Using an explicit  W-witness, 
e.g. 
\be
W_W=\half \eins -\calp_{GHZ} \ ,
\ee
where $\calp_{GHZ}$ is a projector onto a GHZ-vector, we showed in
\cite{abls} that the set of $W\setminus B$ is not of measure zero,
contrary to the pure case \cite{duer}. The idea is to take a state
from the family
\be
\varrho = \frac{1-p}{8}\eins + p \calp_W \ ,
\ee
where $\calp_W$ is the projector onto a pure W-state, and show that
for certain values of the parameter $p$ a ball of states surrounding
$\varrho$ is in $W\setminus B$. This ball is given by
$\tilde\varrho = (1-\kappa)\varrho + \kappa \sigma$, where $\sigma$ is
arbitrary, and the task was to show that there is a finite range
of $\kappa$ such that $\tilde\varrho$ is still contained in
$W\setminus B$.

Another topic that we have studied in the I-Ha programme is
the question whether bound entangled states can be found in
any of the described entangled sets \cite{abls}?
 Our conjecture is that this is not the case, and that bound
entangled states cannot be in $GHZ\setminus W$. 

For small ranks, namely $r(\varrho)\leq 4$,  it is clear that 
bound entangled states even have to be biseparable, i.e. can be
neither in $W\setminus B$ 
 nor in $GHZ\setminus W$: viewing $\varrho$ as a state from the Hilbert space
of $2\otimes 4$ type we can use the result from \cite{pra1} that
any $2\otimes 4$ PPT state with rank $\leq 4$ is separable.

For higher ranks, the idea leading to our conjecture is as follows:
bound entangled states can only be detected by non-decomposable witnesses,
i.e. operators of the form 
\bea
    W_{nd}&=&W_d-\epsilon\one\ , \nonumber \\
    W_{d}&=&P+
\sum_X Q^{T_X}_X\ ,
    \eea
 with $P,Q_X \geq 0$,    $R(P)=K(\varrho)$, $R(Q_X)=K(\varrho^{T_X})$, where   
 $X=A,B,C$. We need to
show that it is always possible to find a W--state $\ket{\phi_W}$
 with $\bra{\phi_W}W_d\ket{\phi_W}\leq 0$, and therefore 
$\bra{\phi_W}W_{nd}\ket{\phi_W}< 0$.
If this is the case, then $W_{nd}$  {\it cannot} be  a GHZ--witness, and thus
$\varrho$ belongs to  
the
$W$--class.

In trying to find $\ket{\phi_W}$ we compared the
number of free parameters
     with the number of equations to solve, which depend on the rank
     of $\varrho$.
It turns out  that there is much freedom, and it is very likely to find
such $\ket{\phi_W}$. We therefore have some evidence for
$\varrho_{BE}\not\in  GHZ \setminus W$.
    
\section{V. Summary}

We have shown that, although the separability problem and the problem
of characterization and classification of positive maps is not yet 
solved, enormous progress has been achieved in the last years.
Nevertheless, in spite of being more than 100 years
old, quantum theory is still extraordinary challenging.

\section{VI. Acknowledgements}

We wish to thank A. Ac\'\i n, 
H. Briegel, W. D\"ur, K. Eckert, J. Eisert, A. Ekert, 
G. Giedke, M. Horodecki, R. Horodecki, 
S. Karnas, M. Ku\'s, D. Loss, C. Macchiavello, 
A. Pittenger, M. Plenio,  J. Samsonowicz,
J. Schliemann, R. Tarrach,  F. Verstraete and G. Vidal 
for discussions  in the context of the
I-Ha programme and at the Gdansk meeting.
This work has been supported  by the DFG (SFB 407 and Schwerpunkt
``Quanteninformationsverarbeitung"), and the ESF PESC Programme on Quantum
Information.


\begin{thebibliography}{99}
\bibitem{foot}{In the mathematical literature some versions of the properties
pointed out in the context of separability were present 
even earlier in terms of cones of positive matrices (see especially 
\cite{posmaps1,posmaps3}, cf. \cite{jami,posmaps2}).}
\end{thebibliography}
\end{document}